\documentclass[conference]{IEEEtran}
\IEEEoverridecommandlockouts
\usepackage{cite}
\usepackage{amsmath,amssymb,amsfonts}
\usepackage{url}
\usepackage{graphicx}
\usepackage{textcomp}
\usepackage{xcolor}
\def\BibTeX{{\rm B\kern-.05em{\sc i\kern-.025em b}\kern-.08em
    T\kern-.1667em\lower.7ex\hbox{E}\kern-.125emX}}
\begin{document}

\title{Introducing A Dark Web Archival Framework}

\author{\IEEEauthorblockN{Justin F. Brunelle, Ryan Farley, Grant Atkins, Trevor Bostic, Marites Hendrix, Zak Zebrowski}
\IEEEauthorblockA{ 
\textit{The MITRE Corporation}\\
Virginia, USA \\
{\{jbrunelle; rfarley; gatkins; tbostic; mhendrix; zaz\}}@mitre.org}
}

\maketitle

\begin{abstract}
We present a framework for web-scale archiving of the dark web. While commonly associated with illicit and illegal activity, the dark web provides a way to privately access web information. This is a valuable and socially beneficial tool to global citizens, such as those wishing to access information while under oppressive political regimes that work to limit information availability. However, little institutional archiving is performed on the dark web (limited to the Archive.is dark web presence, a page-at-a-time archiver). We use surface web tools, techniques, and procedures (TTPs) and adapt them for archiving the dark web. We demonstrate the viability of our framework in a proof-of-concept and narrowly scoped prototype, implemented with the following lightly adapted open source tools: the Brozzler crawler for capture, WARC file for storage, and pywb for replay. Using these tools, we demonstrate the viability of modified surface web archiving TTPs for archiving the dark web.
\end{abstract}


\section*{Funding \& Copyright}
This work was funded by MITRE's IR\&D program; we thank the numerous collaborators that assisted with the award and execution of this research project.
\copyright 2021 The MITRE Corporation. ALL RIGHTS RESERVED. Approved for Public Release; Distribution Unlimited. Case Number 21-1925.

\section{Introduction}
\label{intro}

Capturing the surface web is challenging. 
Archiving the dark web (the portion of the web that requires specialized tools, such as the Tor browser, for access) is harder.  Its anonymity and privacy features make archiving the dark web challenging. While effective for archiving the surface web, institutional archival tools, techniques, and procedures (TTPs) are not effective at archiving the dark web without modification. 

There is no robust architecture or infrastructure for archiving the dark web at scale. Some services, such as archive.is, are available to perform page-at-a-time captures of dark web pages\footnote{The dark web URI-R for archive.is is currently \url{http://archivecaslytosk.onion/}.}. The Internet Archive has a dark web presence\footnote{\url{https://archivebyd3rzt3ehjpm4c3bjkyxv3hjleiytnvxcn7x32psn2kxcuid.onion/}} but does not actively archive the dark web. As such, there is a need to consider a dark web archival infrastructure.

In this paper, we propose a framework for automatically archiving the dark web that uses modifications to proven surface web archiving TTPs. By understanding surface web TTPs and the nuances of the dark web, we propose a framework for archiving the dark web that can produce a historical record of the dark web and preserve dark web content.  
We present a prototype of the framework and discuss how dark web archival TTPs differ from the surface web. We use Memento terminology \cite{nelson:memento:tr} for consistency. We refer to live pages as resources identified by URI-Rs and archived resources as ``mementos'' identified by URI-Ms. We refer to the time at which mementos are archived as the memento datetime.

\section{Introducing The Dark Web}
\label{dark}
The deep web is a term used to refer to portions of the web that are not available to web bots and crawlers. This includes web pages behind authentication, forms, paywalls, and other barriers that human users navigate but crawlers often cannot. The deep web is often hidden from automatic access without specialized tools and \emph{a priori} knowledge. The surface web is the portion of the web that is accessible by crawlers without specialized tools. The dark web is part of the deep web that requires specialized tools for access. 

While the dark web may be typically associated with illicit activity, it is used for a variety of benign purposes. For example, the dark web and its associated access mechanisms allow those under oppressive political regimes to access news and other information that is unavailable through a nationally controlled internet \cite{tor-jardine}. 

The Tor browser is required to access the dark web and offers private or protected browsing by users (a key feature that allows the safety of both disenfranchised web users and those engaged in criminal activity). 
Tor traffic is encrypted throughout its transit through a large network of nodes. Macrina highlights the importance of understanding how Tor works and its non-illicit uses (e.g., for libraries) \cite{tor-macrina2}. 

Dark web sites use the \texttt{.onion} top level domain (TLD) as opposed to \texttt{.com} or \texttt{.org} used on the surface web. 
Universal Resource Identifiers (URIs)\footnote{A URI is a generic form of a Universal Resource Locator (URL).} in the \texttt{.onion} TLD are created based on the public key of the dark web resource that the URI identifies. For example, the New York Times has dark web site (https://www.nytimes3xbfgragh.onion/). Vanity URIs, such as the NYTimes URI, are created by repeatedly generating the URI until the desired string (in this case, the ``nytimes'' portion of the URI which is followed by a series of numbers and letters) happens to get created. If a new public key is issued or the URI is re-generated, the URI for the site will change. 

We observed a set of non-illicit \texttt{.onion} sites from the Real World Onions github page\footnote{\url{https://github.com/alecmuffett/real-world-onion-sites}} and tracked their shifts over time. We analyzed the github changes (in \texttt{master.csv}) which has 15 months of changes to the 128 \texttt{.onions} tracked in the github page. There were 20 updates (i.e., commits) to the github page, most of which are site additions, removals, or format changes. There were 4 commits to account for a change to \texttt{.onion} URIs for 9 sites. Each of the 9 sites only changed once during our 15 month observation period. In other words, over a 15 month period, 9/128 non-illicit URIs shifted (7\% of observed URIs). 

We also analyzed the change logs for a dark wiki\footnote{\url{http://zqktlwi4fecvo6ri.onion/}} (a wiki that similarly tracks \texttt{.onion} URIs). This wiki may contain URIs of dark web sites that contain illicit content, so care should be taken when exploring this dataset. The dark wiki tracks 1,365 sites. From 24 months of change logs on the wiki, we observed 794,234 updates to the wiki with 22 of the updates accounting for a URI shift for 268 sites. This means that our second observed set of \texttt{.onion} URIs shows that 268/1,365 sites (19.6\%) shifted over a 24 month period.

Our observations demonstrate that shifts to URI-Rs of dark web sites may be rare but also unavoidable over time. As a reminder, these are sites that are still ``up'' but are now available from a new URI-R as opposed to a site becoming unavailable or intermittently dereferencable. 
In summary, \texttt{.onion} URIs for a site may shift over time which warrants tracking through a dedicated service.

\section{Surface vs Deep Web Archiving}
\label{archivelandscape}

\begin{figure}[h]
    \centering
    \includegraphics[width=0.4\textwidth]{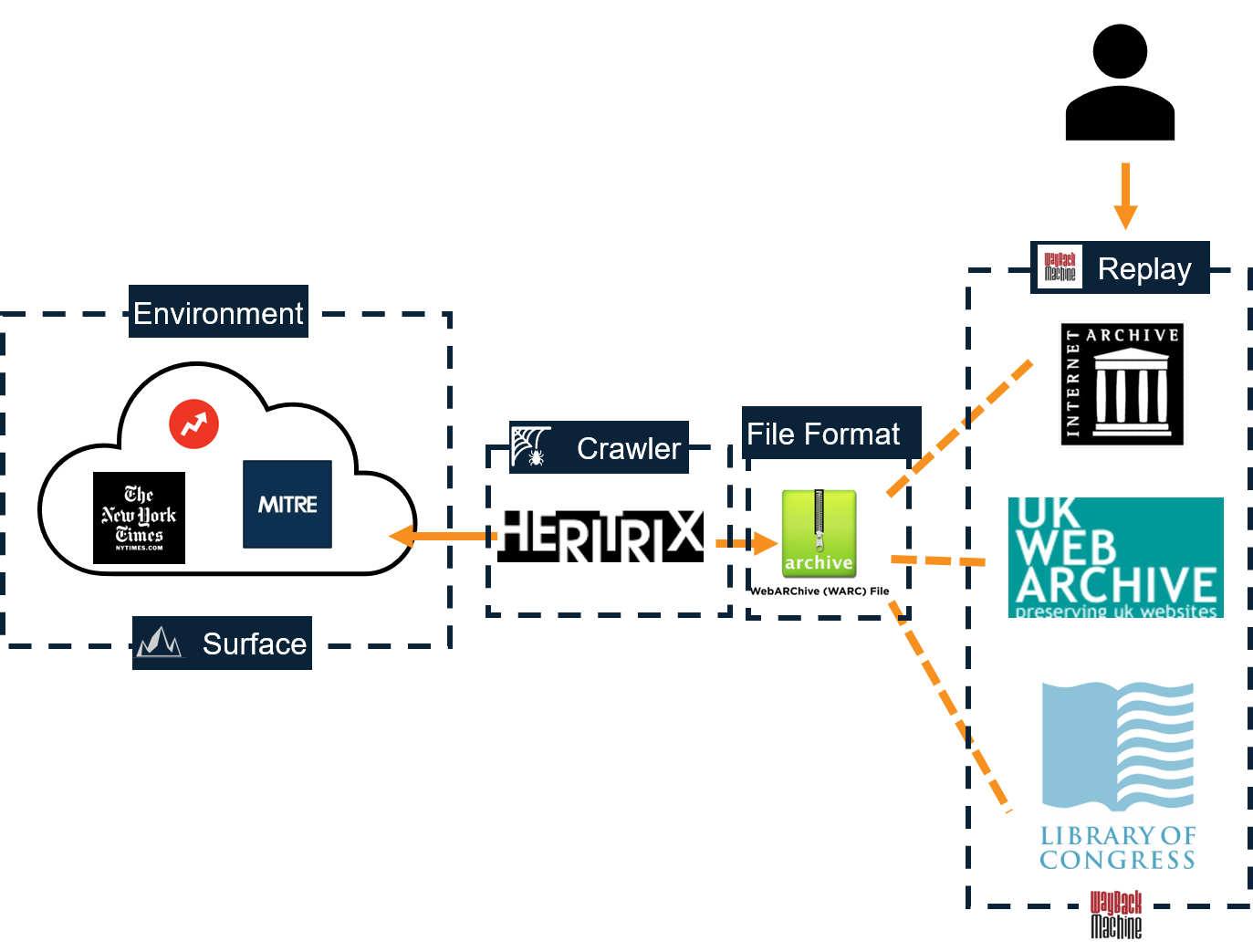}
    \caption{A generic surface web archiving framework includes a crawler, storage, and replay mechanisms.}
    \label{surface}
\end{figure}

Generic web archiving frameworks (Figure \ref{surface}) have three components: the crawler, the storage mechanism, and the replay component. Web archives use crawlers (e.g., Heritrix \cite{heritrix}) to crawl and capture the surface web, store the resulting captures in a WARC file format, and replay the captured content as a memento in their archive (often using the Wayback Machine). 
Archive-It has a traditional framework for archiving the surface web that serves as the basis for ours; it includes a crawler (Brozzler), a replay mechanism (Wayback Machine), and a storage mechanism (WARC). 

Deep and dark web crawling is not new. Madhaven, et al. used a combination of DOM features, form input types, and text to generate queries for HTML forms \cite{deep-web-madhaven}. 
New York University developed a framework for dark web crawling \cite{afrldw} in an effort to improve Darpa's Memex program\footnote{\url{https://www.darpa.mil/about-us/timeline/memex}}. The framework focuses on discovering illicit activity on the dark web through focused crawls. The crawler directs traffic through a Tor proxy but addresses neither shifting \texttt{.onion} URIs nor page replay.


Other efforts leverage ``surface'' web crawlers for discovering illicit content from both surface and dark web forums \cite{dw-sw-crawl}. Gwern, et al. present a model for and results of crawling dark net markets (focused on illicit activity) for analysis \cite{dnmArchives}. Focused crawlers are used to identify extremist or hate speech in dark web forums. Typically, these approaches use NLP to confirm their target captures and involve human-driven capture \cite{dw-crawler} and hashes of content to compare captures without forcing analysts to view illicit content \cite{dw-illicit}. 
While focused dark web crawlers exist, none address the needs of a dark-web-scale \emph{archival} framework, such as archival-quality capture, storage, and replay of dark web pages and embedded resources. 

The dark web is difficult to crawl \cite{dw-crawl-challenges} and shares some archival challenges with the surface web (e.g., scale, login walls) but others are unique to the dark web. We compare some of the challenges with crawling and archiving the surface and dark webs in Table \ref{challengetable}. 

\begin{table}[]
\begin{tabular}{ p{1.4cm} | p{3.6cm} | p{2.6cm} }
\textbf{Challenge} & \textbf{Surface Web} & \textbf{Dark Web} \\
\hline
\hline
URI canonicalization & Parameters and ports must be resolved during replay and analysis & \texttt{.onion}s shift; canonicalization must occur during crawl, replay, and analysis\\
\hline
Crawl Topics & Crawl everything, check for malicious JavaScript; contraband is detected post-crawl & Crawl only what is allowed by policy, need to check for malicious JavaScript and illicit material\\
\hline
Crawl method & Open source tools for web-scale collection maintained by large institutions (e.g., Internet Archive) & Tools exist but are not suitable for the entire web archiving framework\\
\hline
Storage method & Often uses headless crawling to capture responses and standard metadata for WARCs; occasionally uses headful crawling & Must use Tor to access \texttt{*.onion}; metadata captured must be standardized using WARC fields\\
\hline
JavaScript & Ajax and other dynamics hide content from headless crawlers and is mostly addressed with headful crawlers (with the exception of interactions) \cite{brunelleDiss} & JavaScript is rarely used due to security and privacy vulnerabilities that can be introduced\\
\hline
Login & \multicolumn{2}{p{6cm}}{Since this content is part of the deep web, only top level crawled (e.g., facebook.com)}\\
\hline
Robots.txt & \multicolumn{2}{p{6cm}}{Crawler politeness configurations determine adherence to robots.txt}\\
\hline
Capture Quality & Archivists and crawl owners typically consider captures of the main HTML and important embedded resources as ``good enough'' even if CSS and other minor resources are missed, whereas common web users consider these resources essential to capture quality and completeness \cite{damageIJDL} & The same challenges from the surface web apply to the dark web, but metadata is much more important for analysts to understand what was captured and how it was archived\\
\end{tabular}
\caption{While the surface web and dark web share some archival challenges, the dark web presents some unique challenges to a web archiving ecosystem.}
\label{challengetable}
\end{table}

\section{A Dark Web Archiving Framework}
\label{framework}
\begin{figure}[h]
    \centering
    \includegraphics[width=0.48\textwidth]{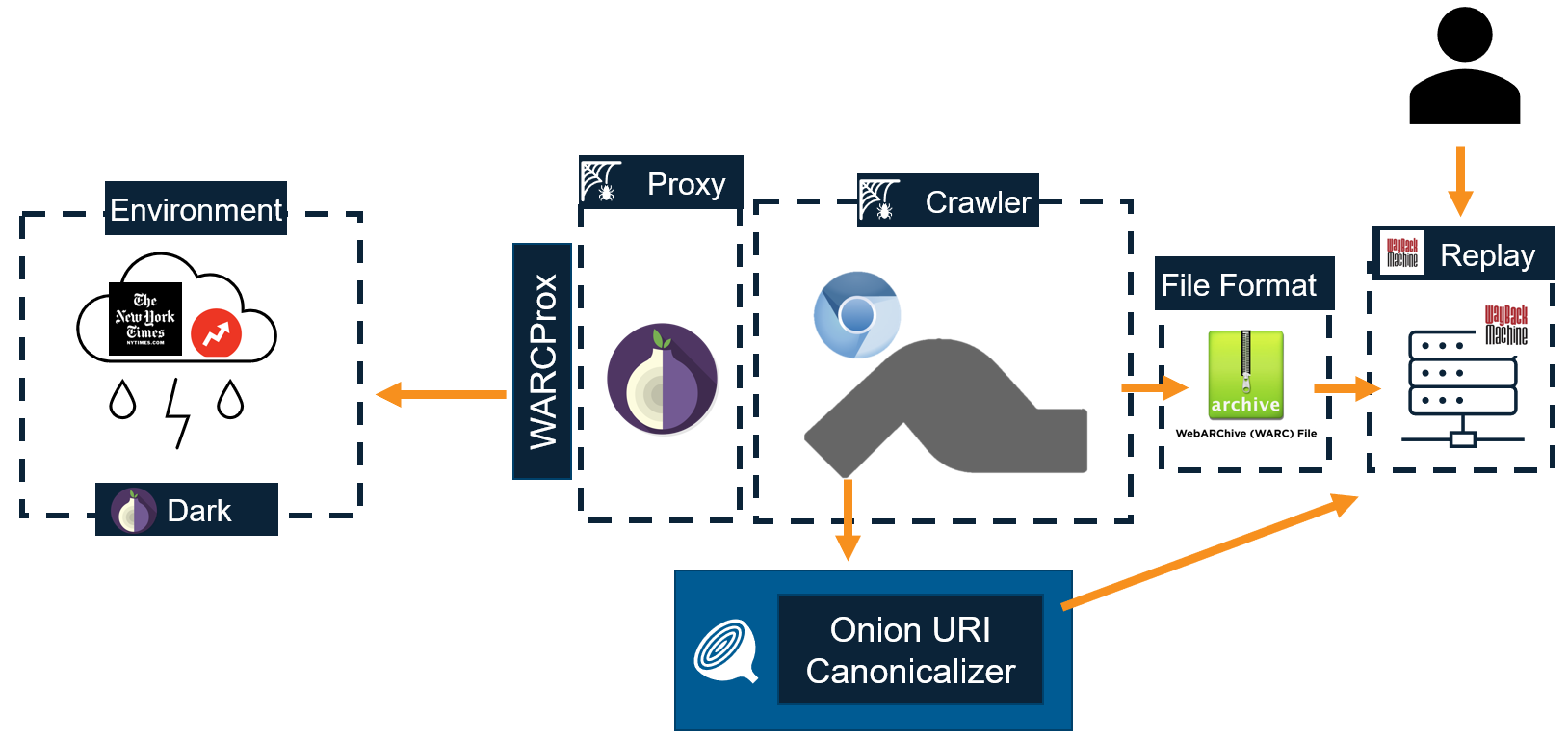}
    \caption{The generic dark web archiving framework is adapted from the framework for archiving the surface web. Notable changes are the Tor proxy to enable dark web access and the \texttt{.onion} URI canonicalization service.}
    \label{darkf}
\end{figure}

Our framework for archiving the dark web replicates the components of the framework for archiving the surface web (presented in Figure \ref{surface})
. Our framework (Figure \ref{darkf}) includes an onion canonicalizer (Section \ref{frameworkindex}), a crawler that archives the dark web (Section \ref{frameworkcrawler}), stores captures as WARCs (Section \ref{frameworkstorage}), and replays mementos using a Wayback Machine (Section \ref{frameworkreplay}).

\subsection{The Onion Canonicalizer}
\label{frameworkindex}

Because the URI-Rs of dark web sites may shift over time, any dark web archival framework will need to canonicalize the \texttt{.onion} addresses over time. This goes beyond URI canonicalization used by surface web archival institutions (e.g., removing fragment identifiers, port :80) to understanding that URIs will shift and should be treated as equivalent. 

Our onion canonicalizer tracks and recognizes new sites, allows the lookup of a \texttt{.onion}'s prior URI-Rs and a prior URI-R's current URI-R, and builds timemaps of a site that includes all of its observed URI-Rs, even if they shift over time. 
The onion canonicalizer records the URI-R (e.g., \url{https://bfnews3u2ox4m4ty.onion}), source (e.g., real-onions github), and alias (e.g., ``Buzzfeed News'') of each dark web crawl target. As the URI-Rs of the pages change, the onion canonicalizer tracks the shifts to build a list of URI-Rs for a site and the time at which the URI-R was observed. This means that the URI-R is tracked using a 4-tuple: \textless URI-R, source, alias, observation time \textgreater. 

Using the 4-tuple, the onion canonicalizer can track URI-R changes over time. If aliases differ between sources but the URI-Rs are the same, the canonicalizer will recognize that the differing aliases define the same site since they point to the same URI-R. The canonicalizer will keep track of this association as it monitors URI-Rs. The onion canonicalizer will track shifts as long as the URI-R and source/alias do not change simultaneously. In our future work, we will automatically detect these collisions or disconnects and elevate them to a human administrator for resolution.

The onion canonicalizer will create a timeline of URI-R, timestamp pairs. The crawler consults the onion canonicalizer as it adds URI-Rs to its crawl frontier and as it prepares to crawl a target to ensure that it is attempting to crawl only the current \texttt{.onion} URI-R. Users can select the most appropriate URI-R for playback based on the desired memento-datetime. 
The onion canonicalizer has an API that allows a URI-R lookup and returns a JSON representation of the requested data. It allows a lookup of the current URI-R based on a prior URI-R, 
a list of each prior URI-R and its observed timestamp based on a current or prior URI-R (i.e., a site's timemap of URI-Rs), and 
a prior URI-R given a target observation timestamp and a current URI-R.

The onion canonicalizer is the only custom code in our framework prototype. This demonstrates an addition to our surface web TTPs to make dark web archiving successful.

\subsection{The Crawler}
\label{frameworkcrawler}
We selected the Brozzler crawler due to its native storage in WARC format, ability to easily interface with a proxy, and ability to perform high-fidelity headful crawling. We opt for high-quality crawling to ensure that JavaScript is executed in the instances that JavaScript is used by our crawl targets. We send Brozzler's traffic through a Tor proxy, allowing it to access the dark web with minimal intrusion and modification of the Warcprox code.

The only code modification we made to Brozzler was to allow it to interface with the onion canonicalizer. When Brozzler references it's own crawler frontier (e.g., retrieving the next URI-R it will crawl), we check if the hostname of the URI-R in Brozzler exists in our onion canonicalizer. The onion canonicalizer provides Brozzler with the current \texttt{.onion} URI-R to target. If the URI-R has shifted, Brozzler will update its crawl target with the new hostname. If the URI-R has not shifted, Brozzler continues its crawl. Brozzler checks the current hostname of its crawl targets using the onion canonicalizer to ensure that Brozzler is crawling the correct target, even if the \texttt{.onion} address of the site has shifted since the URI-R was added to the frontier.


\subsection{The Storage}
\label{frameworkstorage}

We use the WARC file format in our dark web archival framework and prototype. This simplifies the integration between the crawl and replay components of the framework. However, to provide additional information regarding onions that shift, we add a metadata field that designates the first observed URI-R of a site into the record of the memento. We prototyped the addition of the metadata field (which is filled by the crawler) to ensure that the WARC records are not invalidated, malformed, and are properly replayed in standard open source Wayback Machine tools. 

\subsection{The Replay}
\label{frameworkreplay}

We use pywb as our replay mechanism. The modified WARCs produced by our crawler are replayable in pywb without any modification. We recognize that pywb will need modification to interface with the onion indexing service to allow seamless integration of dark web mementos with shifted URI-Rs by pywb. In other words, if pywb is replaying a memento with a URI-R that has shifted, pywb will need to be modified to perform lookups of the URI-M to allow users to natively navigate through the archive. 
Our current pywb implementation leaves the lookup of shifted URI-Rs as a task for the human user. Changing this functionality to remove human interference is a future goal.

\subsection{The Implementation}
\label{implementation}

We deployed our framework on a virtual machine (VM) with a Tor proxy to handle outbound traffic. Brozzler used the proxy to visit both dark and surface web resources and is configured to avoid spidering beyond pre-configured addresses to avoid the accidental capture of illicit material. 
We crawled and replayed two sites over 31 days (during which the observed URI-Rs never shifted). This resulted in the capture of 75,956 mementos and 16 GB of WARCs generated. 
We also observed that 8/78 (10.2\%) of observed sites from our dataset use JavaScript. Despite the fact that using JavaScript on the dark web can eliminate any anonymity of the user, its presence justifies our use of a high-fidelity headful web crawler despite its slower crawl speed than its headless couterparts \cite{brunelleDiss}. 


\section{Conclusions}
We present a framework for archiving the dark web as a first step in applying surface web archiving TTPs to the dark web. We hope that this paper motivates further research from the web archiving community. While illicit  activity occurs on the dark web, there are many socially beneficial uses of the dark web; yet it receives little focus from the web archiving community. Because the dark web is not actively archived at scale, historians and future analysts will be unable to study how content is delivered to users. Our framework can serve as the web-scale counterpart to the Archive.is dark web page-at-a-time archival service. 
We discuss the challenges with archiving the dark web, compare them to the surface web, and present a framework that addresses these challenges (Table \ref{solutionstable}). To demonstrate our framework's viability, we implemented a proof-of-concept prototype to archive two target dark web resources over a period of 31 days, collecting 75,956 mementos which totalled 16 GB of WARCs. 
Our future work will focus on modifying pywb to interface with our onion indexing service. We will also measure the performance tradeoffs of our design choices (such as a headful vs headless crawler \cite{crawlSlowly}).

\begin{table}[]
\begin{tabular}{ p{1.4cm} | p{6.6cm} }
\textbf{Challenge} & \textbf{Framework Solution} \\
\hline
\hline
URI canonicalization & Create an onion indexing service that monitors and records the URI-Rs of dark web archival targets. Archive the dark web resource and record the observations of its URI-Rs at crawl time in the WARC; allow the replay service to look up the correct URI-R during replay to construct the appropriate URI-M. \\
\hline
Crawl topics & Adopt crawling solutions with configurable scope and crawl policies to avoid archiving undesirable material.\\
\hline
Crawl method & Leverage a crawler that captures WARCs and adapt it to provide the relevant metadata associated with the crawl and capture. \\
\hline
Storage method & Use WARCs to store captured content and provide URI canonicalization metadata for future replay. Capture a single URI-R in a single WARC with adapted metadata. \\
\hline
JavaScript & Dark web sites often avoid JavaScript, but some still use it (e.g., NY Times, Buzzfeed). Use headful browsing tools to capture JavaScript-driven representations and mitigate the effect of deferred representations on mementos. \\
\hline
Login & Archive the base page, as done on the surface web.\\
\hline
Robots.txt & Obey robots, as done on the surface web.\\
\hline
Capture Quality & Focus on high-fidelity crawling through the use of headful crawling approaches and ensure high-fidelity playback.\\
\end{tabular}
\caption{Our proposed framework addresses a subset of the challenges with dark web archiving.}
\label{solutionstable}
\end{table}




\end{document}